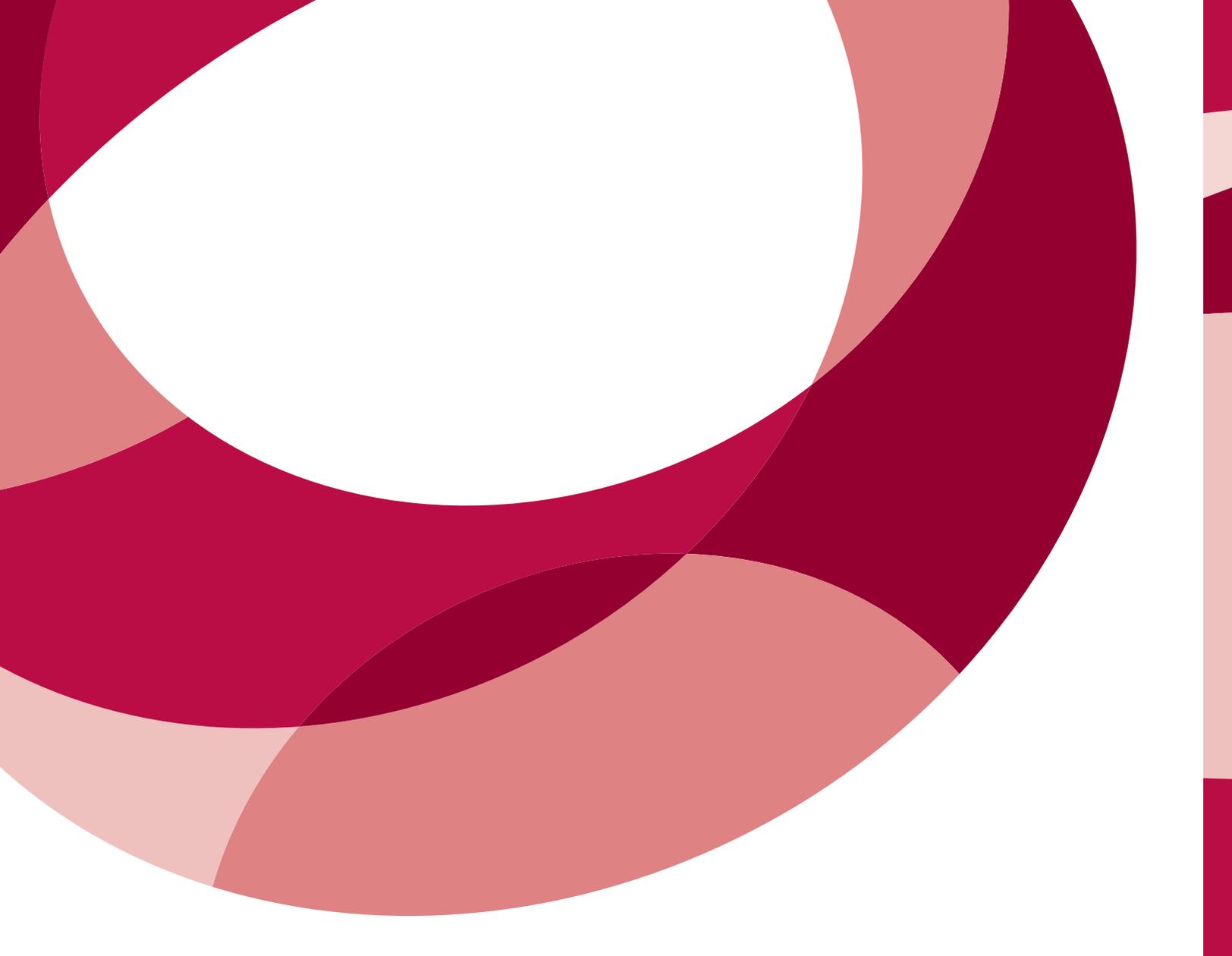

# Promoting Strategic Research on Inclusive Access to Rich Online Content and Services

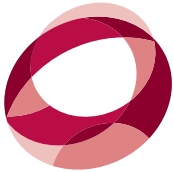

CCC
Computing Community Consortium
Catalyst

This material is based upon work supported by the National Science Foundation under Grant No. (1136993). Any opinions, findings, and conclusions or recommendations expressed in this material are those of the author(s) and do not necessarily reflect the views of the National Science Foundation.

# Promoting Strategic Research on Inclusive Access to Rich Online Content and Services

Shaun Kane, Richard Ladner, and Clayton Lewis

Sponsored by

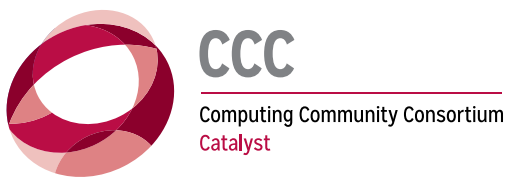
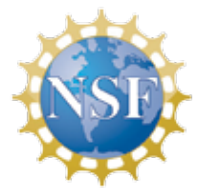

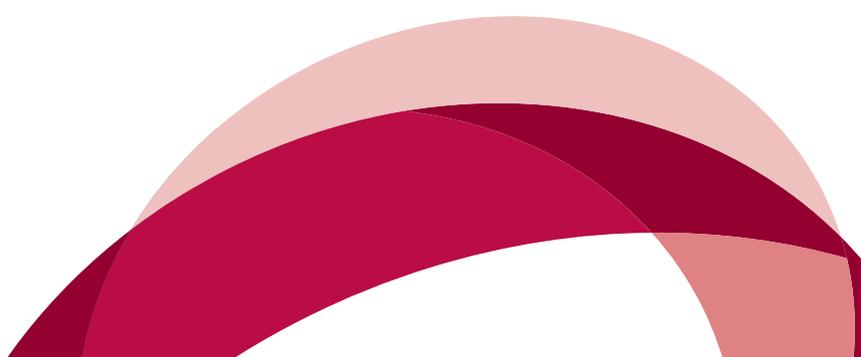

## Executive Summary

Access to content and services online is increasingly important for everyone, including people with disabilities. National commitments, including the Americans with Disabilities Act, and international resolutions, including the United Nations Declaration of the Rights of Persons with Disabilities, call for work to ensure that people with disabilities can participate fully in the online world. Gains in education, employment and health, as well as in civic engagement, social participation, and personal independence will follow from enhanced inclusion online. Research in many areas of computer science, including recognition technology, natural language processing, personalization, software architecture, and others, is needed to secure these benefits. Organizing this research calls for partnerships among academic researchers, federal agencies, and commercial organizations, as well as effective division of labor and cooperation between computer scientists, behavioral scientists, advocacy groups, and consumers.

## 1. Introduction

### 1.1 Need, Opportunity, and Impact

Access to content and services online is increasingly important for everyone, including people with disabilities. Since the launch of the World Wide Web, in the early 1990's, online access has become more widely used in many areas of life. In health, the Pew Internet trust reports that more than 70% of consumers look for health information online[1]. In education, sites like Khan Academy, as well as providers in higher education like Coursera and EdX, offer content at all levels, from kindergarten through postgraduate education; almost 90% of public universities offer online courses[2]. Almost 90% of job seekers report looking for opportunities online[3], and almost 90% of recruiters report hiring through online social media[4]. More than half of US consumers bank online[5]. Online shopping is growing rapidly[6]. Consumption of news online is growing rapidly[7]. Games, films, and other forms of entertainment are increasingly consumed online[8]. Increasingly, online engagement happens on smart phones, with nearly two-thirds of consumers having these devices[9].

Inclusion of people with disabilities in this online world is an important priority, internationally and nationally. Article 9 of The UN Convention of the Rights of Persons with Disabilities, adopted in 2006 and ratified by 160 countries, calls for "appropriate measures to ensure to persons with disabilities access, on an equal basis with others, to the physical environment, to transportation, *to information and communications, including information and communications technologies and systems*, and to other facilities and services open or provided to the public[10] (emphasis added)." In the US, the Declaration of the Rights of People with Cognitive Disabilities[11] (see also Blanck, 2014) affirms that "Access to comprehensible information and usable communication technologies is necessary for all people in our society, particularly for people with cognitive disabilities, to promote self-determination and to engage meaningfully in major aspects of life such as education, health promotion, employment, recreation, and civic participation." The Americans with Disabilities Act of 1990 commits the USA to providing that people with disabilities have the same opportunities as everyone else to participate in the mainstream of American life. Online technology makes important contributions to meeting that commitment. Continued progress towards free and full access to content and services online will serve additional interests, as well:

- Government agencies need better ways of communicating with the public, including people with disabilities. Access to information is essential in preparing for and responding to natural disasters. But experience with hurricanes Katrina and Sandy showed that our ability to communicate effectively with people with disabilities is inadequate. Enhanced accessibility of online media is important in addressing this need.

- Government agencies need better, less costly ways of meeting accessibility commitments in law. Section 508 of the Rehabilitation Act requires federal agencies to acquire accessible technology, including online tools and resources. Meeting this requirement today requires careful training of many staff, and while progress is being made in increasing the effectiveness of this training, many agency websites and communication programs are not yet fully accessible.

- Inclusive access online will increase employment for people with disabilities. Today the employment rate for people with disabilities is only slightly over 17%, as compared to more than 64% for the general population[12]. As more opportunities for online work become available, making these accessible to people with disabilities will increase employment, decrease the cost of social services, and increase economic productivity.

---

[1] http://www.pewinternet.org/fact-sheets/health-fact-sheet/
[2] http://www.pewsocialtrends.org/2011/08/28/the-digital-revolution-and-higher-education/
[3] http://www.pewinternet.org/2015/11/19/searching-for-work-in-the-digital-era/
[4] http://socialmeep.com/infographic-the-social-recruiting-pocket-guide/
[5] http://www.pewinternet.org/2013/08/07/51-of-u-s-adults-bank-online/
[6] http://www.census.gov/newsroom/press-releases/2014/cb14-102.html
[7] https://transition.fcc.gov/osp/inc-report/INoC-20-News-Consumption.pdf
[8] https://www.ntia.doc.gov/files/ntia/publications/exploring_the_digital_nation_-_americas_emerging_online_experience.pdf
[9] http://www.pewinternet.org/2015/04/01/us-smartphone-use-in-2015/
[10] http://www.un.org/disabilities/convention/conventionfull.shtml
[11] http://www.colemaninstitute.org/declaration

Online delivery of information is already eliminating or easing barriers to access for many people with disabilities. Blind people cannot use traditional print media, without slow and costly conversion to Braille or spoken form. But text in machine-readable form can be made accessible quickly and at virtually no cost by text to speech technology, so (for example) almost any newspaper in the world can now be read online by blind citizens. Similarly, online delivery of video can incorporate captions, allowing access for deaf and hard of hearing consumers. Online commerce can be easier for people for whom traveling to a store, and/or interacting with store personnel, is difficult. Online educational content can be more accessible than in-classroom experiences. See Harper and Yesilada (2008) for discussion of progress along these lines.

The time is ripe for increased research investment to build on and extend these successes, for three reasons. First, the aging population means that the impact of inclusion will increase. Disability is strongly associated with age, with more than a third of people over 65 having a disability that affects their ability to perform tasks of daily living. The number of people over that age, and hence the number projected to be living with disabilities, is increasing[13]. Accessible online technology will be important in supporting the well being of this aging population. Second, there are barriers to inclusion that are increasing, rather than decreasing, and we need to act now to reverse this trend. New forms of information delivery, such as infographics, online maps, and interactive simulations, are increasingly being used, and we lack effective means of making these mediums accessible to people who cannot see well. Finally, advances in technology are creating opportunities to reduce existing barriers. For example, complicated text is difficult to understand for many people, but advances in technology for processing natural language may make it possible to address this challenge. Advances in automatic recognition technology and machine learning may make it possible to convey the emotional content, as well as literal text, in providing captions for video content. Further, progress in software architecture may make it possible to greatly reduce the cost of accessible technology, while making it more widely available.

---

[12] http://www.bls.gov/news.release/pdf/disabl.pdf
[13] http://www.aoa.acl.gov/Aging_Statistics/Profile/2014/4.aspx

Here are illustrations of the potential impact of progress in this work:

- A job seeker can find job opportunities and complete job applications online, regardless of disability. The presentation of information, and how they interact with it, is automatically tailored to their capabilities.

- A blind student can choose from a wide array of advanced technical courses, available online, knowing that all the information presented via diagrams and interactive simulations will be accessible.

- A consumer with limited hand movement can get health information online without needing to use a mouse or touchpad.

- A deaf film buff enjoys easy access not only to what is being said, but also to which character is speaking, and their emotional tone.

- A deaf person can communicate with a hearing colleague via speech to text, knowing that their speech will be accurately interpreted.

- A blind television viewer finds that audio descriptions are readily available for any desired content.

- A student with dyslexia can ask questions to extract desired information from a reference book, without having to read extended text.

- A person experiencing age-related cognitive decline can read about a Social Security benefit in language tailored to their vocabulary.

Progress will lead to impact beyond these important results. In developing collaboration between technology researchers and researchers and developers focused on the needs of people with disabilities, we will see breakthroughs in technology for everyone.

**1.2 Drawing a Roadmap for Progress**

With sponsorship by the Community Computing Consortium, a workshop was convened in Washington, DC to discuss research opportunities in inclusion online. Meeting over two days, September 25-26, a group of 39 computer scientists, representatives of disability advocacy organizations, people from industry, and nine federal employees discussed opportunities in breakout groups. They focused on six areas, automatic description of image and video content, online support for deaf people, access to textual content for people with language and learning disabilities, inclusive design of games and simulations, access to large quantitative datasets, maps and 3D printing, and software architecture for configurability.

Figure 1 summarizes the findings from the workshop, highlighting the logic for addressing the needs and opportunities. The left side of the figure shows the areas of research with most promise, both as sources of new solutions for people with disabilities, and as opportunities for progress from focusing on the needs of inclusion. The middle of the figure shows the key actions needed to reach the key goal of inclusive access.

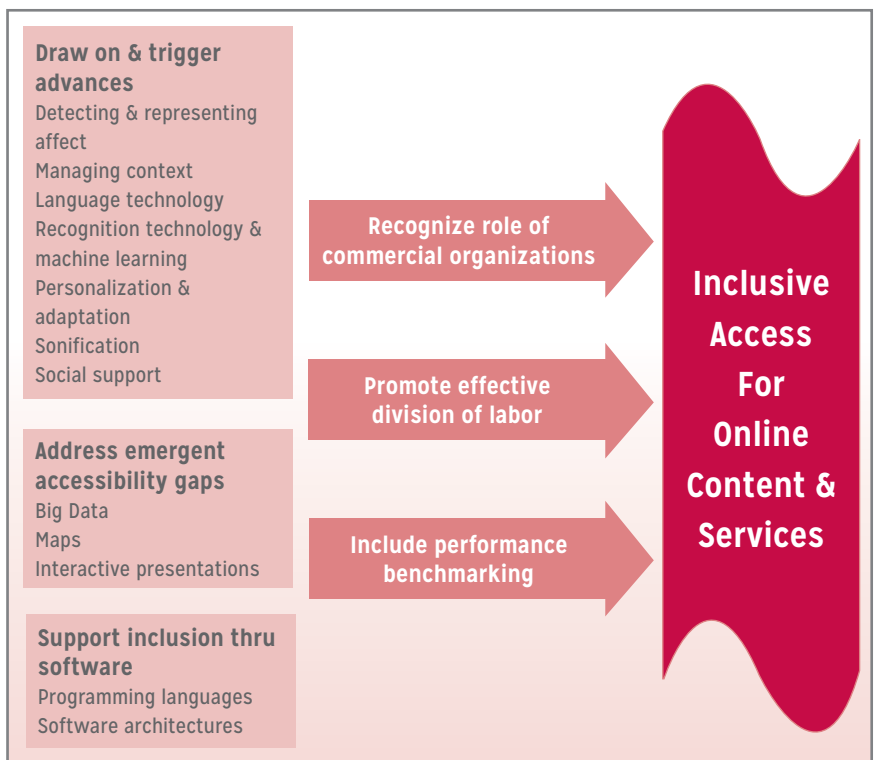

Figure 1: Roadmap

## 2. Research Opportunities

Workshop participants identified two broad classes of research opportunities. First, there are new developments in science and technology that can help address existing barriers. We can and should draw on scientific and technical advances in other fields to better support inclusive design online. Participants suggested that there is potential for major advances in inclusion if researchers and developers working on accessibility can work cooperatively with researchers who are developing new technologies for other purposes, as well as with scientists seeking fundamental understanding of relevant problems. Second, some opportunities arise because new technologies for delivering content and services create new barriers. For example, the increasing use of games and simulations, for education as well as for entertainment, creates a need for research on how to support the use of these media by people with disabilities.

**2.1 Draw On and Trigger Advances in Other Fields**

**Research theme: Detecting and representing affect.** Captioning is crucial in supporting access to video content for deaf and hard of hearing people. But current captioning technology seeks only to convey the literal content of speech, and not the affect that spoken language vividly conveys. Research advances in assessing affect in spoken communication may be exploited, along with the development of new means of presenting captions or caption-like material, to support much fuller access to the meaning of speech by deaf and hard of hearing people.

**Research theme: Managing attention in multi-stream communication**. When missing one sense, other senses can compensate to some degree. However, in using one sense, like vision, one may need multiple visual streams of information to get the same information as when having both vision and hearing. For example, a person who is deaf must consume both the captions and the video when viewing a movie. Captions and audio descriptions (auxiliary content that conveys what is happening in video, for blind consumers) require consumers to process multiple streams of information at the same time. Similar challenges occur for deaf students processing a lecture, when sign language interpretation accompanies slides. Research on attention management in multitasking, suggests that people may be able to process multiple streams of content simultaneously, under some conditions (Guerreiro and Gonçalves, 2015). This work has the potential to lead to tools with revolutionary benefits for everyone, including people with disabilities.

**Research theme: Managing context in interpreting communication.** The meaning of information is nearly always conditioned by context. For example, the meaning of an infographic depends on the meaning of supporting text, and the meaning of text depends on the meaning of an adjoining diagram. An appropriate description of an image, provided to give access to a blind reader, depends on surrounding text. Approaches are emerging in computer vision research that can describe images out of context; how can these approaches be adapted to take advantage of context to automatically create meaningful alternative text for images?

**Research theme: Language technology.** A number of developments in automated natural language processing have potential for broadening access to information online. Improved technology for generating natural language descriptions may improve access to large quantitative datasets for blind consumers. Advances in question answering (Tesauro et al., 2013; Guha et al., 2015) may support broader access by allowing consumers to ask questions about the meaning of information presentations of all kinds, rather than having to extract meaning directly. The ability to extract meaning from complex texts, and then re-present the meaning in simpler, more easily understood form; automatic generation of an outline or graphic organizer for lengthy text; a "dictionary" function that could determine the meaning of a word in context rather than just list of all its possible meanings; and automatically breaking up text into phrases for a more understandable visual presentation all may enhance access for consumers with cognitive and language limitations.

**Research theme: Recognition technology and machine learning.** Emerging approaches such as "deep learning" are enhancing our ability to automatically determine the content of a variety of signals, including images, video, and speech. These technologies can be shaped so as to provide alternate descriptions of images, audio descriptions of video content, and transcriptions of nonstandard speech, including deaf speech. Research in this area should support the VIRAD, the Visual Impairment Research Agenda for Description[14], being developed jointly by the Smith Kettlewell Institute and the National Center for Accessible Media.

A more immediate application, that should be in reach of current technology, but has not been realized, is automatic detection of document structures such as headings. Currently many documents are inaccessible to blind readers because their structures are not detectable by screen reader tools. Avoiding this requires training anyone who creates online content, which is both costly and often ineffective. Automatic recognition of structure would reduce cost and increase effectiveness.

**Research theme: Personalization and adaptation.** Different people, with different functional capabilities, require information to be presented differently. Today individuals must spend time and effort configuring systems to meet their needs, and often public systems, such as information kiosks or ticket machines, cannot be configured appropriately. Technology is emerging that supports auto-personalization, in which an online specification of user needs and preferences is used to configure any system they use, automatically[15].

A related development is the ability to determine desirable configuration automatically, from observing user behavior. This has been demonstrated in some situations, for example keyboard settings (Trewin). Added capability may be possible for setting font size and text formatting.

**Research theme: Sonification.** Our ability to produce synthetic sound, with vivid spatial localization, is increasing. This technology can support a wide range of new ways of presenting information to users who cannot see well. This technology is being used to present mathematical structures, like curves. Extensions may be possible for presenting more complex structures.

**Research theme: Tactile displays.** Spatial patterns detectable by touch have long been used to display Braille. More generally, tactile displays can support users who cannot see or hear, and may be able to do a better job of presenting spatial relationships than sonification. But these displays are expensive, and have limited size and resolution. Dynamic tactile displays capable of displaying larger patterns are possible, but too expensive for general use. New technologies based on MEMS, nanotechnology, and smart materials are emerging, and increased investment to accelerate their development is appropriate. Research is also expanding in providing tactile feedback to flat touchscreens.

**Research theme: Social support for technology use and development.** Use of new technology requires learning, and this learning is widely supported today by online social processes, as seen in sites like Stack Overflow, or community forums associated with new products. These social technologies, if they are themselves accessible, have important potential for people with disabilities if they make it possible for people with disabilities to share information about their particular needs. Today, people with disabilities too often must depend on people without disabilities to discern their needs and respond to them. Enhanced social media support would empower more people with disabilities to learn about and adopt appropriate technology, and to develop and share technology that meets their needs.

There are also social barriers to be overcome. Disability often carries a stigma with it, as may use of assistive technology. People are reluctant to ask for help or use any supports provided if they are seen in this light. More universal design – merging access technology into mainstream technology (like pervasive availability of zooming and speech input on mobile devices) – will help with this. Further, more research on the affective side of disabilities and accessibility technologies would also be worthwhile.

---

[14] http://www.ski.org/project/visual-impairment-research-agenda-description-virad
[15] www.GPII.net

**2.2 Address Emergent Accessibility Gaps**

**Research theme: Presenting large quantitative datasets.** Many entities, including the press as well as public agencies, are publishing large quantities of data online, with visual access provided by charts or graphs, often of sophisticated design. For example, the Oceans of Data Institute has done a lot of work on the use of large datasets in education, emphasizing visual interfaces (http://oceansofdata.org/our-work/visualizing-oceans-data-educational-interface-design). Unfortunately, these visual presentations are not accessible to blind people, depriving them of the benefits they provide, including opportunities to understand and participate in public discourse concerning the data. Research is needed on a number of fronts to address this challenge, including:

- Large tactile displays
- Automatic generation of linguistic descriptions, from data and from information visualizations
- Question-answering interfaces for quantitative datasets

**Research theme: Maps.** It is now easy to provide maps in online presentations of all kinds, but we lack broadly applicable means of making this information accessible to blind people. Research on this challenge should include:

- Understanding the diverse purposes maps serve
- Devising accessible presentations appropriate to particular purposes of maps
- Developing inexpensive tactile displays capable of displaying map data, including via 3D printing.

**Research theme: Inclusive design of interactive presentations.** It is now easy to provide interactive simulations of complex systems, such as electronic circuits or business operations, and these are commonly used in education. Similar technology powers interactive games that are increasingly used in education, and to build public engagement (see for example https://www.challenge.gov/challenge-gov-celebrates-five-years-of-open-innovation/) as well being an important form of entertainment. But these highly interactive experiences pose serious accessibility challenges, including for blind users and users with motor limitations that make a mouse or touchscreen unusable.

Research is needed to address these accessibility challenges, including:

- Developing representations capable of capturing the content of an interactive simulation or game in forms not tied to visual presentation.
- Exploring whether such representations can be made fully amodal, that is, not tied to any specific sense modality, but being adaptable to any modality.
- Developing software structures that make it possible to substitute alternative modes of interaction, such as keyboard control for touchscreen, while not disrupting other aspects of a system.
- Making it possible for learners or players with and without disabilities to participate in the same learning or play activity together.

**2.3 Support Inclusion Through Software**

A broader need that emerged in this survey of research opportunities is for software structures that make accessibility enhancements easier to implement. Today, adding accessibility features, such as support for a tactile display, requires substantial implementation effort, especially if the need was not identified and planned for during initial design. As the pace of innovation increases, this is no longer acceptable. Rather, software structures are needed that make it easier to substitute different representations for the same information, and different ways of interacting with it.

Progress in this area will pay dividends beyond advances in inclusion. Developing software products of all kinds in the current environment, in which display sizes and interaction modes (pointing device vs touchscreen) differ, is challenging and expensive. "Responsive" design methods that aim to support varying screen size help a little, but are themselves costly, and do not address other aspects of variation. As new interaction techniques and device categories proliferate, software structures that provide more flexibility will be very valuable.

Research on programming languages and software architectures that aim to address this challenge includes the KORZ, NewSpeak, and Fluid projects. Enhanced support for this line of work is needed.

This software research can provide a crucial added benefit: it can make it easier for people with disabilities themselves, and their caregivers, to create useful technology that meets individual needs. As mentioned earlier, people with disabilities are too often dependent on technologists without disabilities to identify and address their needs. Making software development tools themselves more accessible, and easier to use by people with less technical skill, is an important research challenge. Research in end user programming, including programming by demonstration and by example, has much to contribute here. See Ladner (2015) for a discussion of the critical importance of empowering people with disabilities to shape technology for their own use.

This research will create new opportunities in society at large. Developments that make information technology easier to create, and easier to shape to particular needs, will greatly enhance the value of the technology. Individual consumers, including small business owners, will be able to address their particular needs, rather than relying on expensive and often unavailable support from technical experts. Making the tools accessible, as well as comprehensible, will enhance economic opportunity for people with disabilities.

## 3. How Can We Grasp These Research Opportunities?

For research to meet the goals of inclusion online, the contributions of individuals with different skills, and organizations with different roles and capabilities, need to be coordinated. Here are some of the considerations in doing this.

**3.1 Recognize the Role of Commercial Organizations**

In promoting research in response to the opportunities described here we should seek to involve technology industry directly. Too often, the research pipeline that starts in academic research does not impact products and services available to consumers and employers, or does so only after a very long delay. For example, research on automatic identification of document structure needs to be embodied in the content creation tools used by federal agencies to create online content. This will happen much more quickly if the companies that create those tools participate in the research.

A public-private partnership may be a way to create the needed collaboration. Federal agencies could broker funding for cooperating teams of researchers and developers in academe and industry.

**3.2 Promote Effective Division of Labor**

Many of the technical and scientific advances that have promise for enhancing inclusion online are being made by people with little or no understanding of the opportunities to serve the needs of people with disabilities. Accordingly, they may have little ability to appropriately evaluate new technology for this purpose, both because of limited contact with people with disabilities, and hence limited understanding of their needs, and because of lack of knowledge of the behavioral science methods needed for evaluation. At the same time, people well informed about these needs, and these behavioral science methods, may have little understanding of the scientific and technical developments to be exploited.

To address these limitations research sponsors should be proactive in assembling the teams needed to carry on this work. Supporting assessment programs separately from technology development programs, but requiring cooperation between the two, could be valuable. This would allow scientists and technologists to pursue work useful for inclusion, without requiring them to recruit study participants and carry out behavioral assessments themselves.

### 3.3 Include Performance Benchmarking

Relatedly, research sponsors should develop benchmarks that will allow the field to determine when progress is being made in meeting inclusion goals, as is essential for rational resource allocation (and for the conduct of research, itself.) Creating benchmarks that address specific federal agency needs, such as delivery of health information, educational materials, published datasets, and public alerts can accelerate progress.

For example, one or more large quantitative datasets, now being published with infographics, could be used as benchmarks. Assessments could be developed to measure the extent to which blind consumers can extract needed meaning from proposed, more accessible presentations. Recruitment of test participants, and the conduct of the assessments, would be provided for any participating, funded project. Thus technology researchers can propose promising work without undertaking the assessment work themselves. Because benchmarking is done consistently for all such projects, the benefits from alternative technical approaches can be compared.

This benchmarking approach will allow research sponsors to balance investment in particular promising technologies with the opportunity for investigators to demonstrate potential for new approaches, using benchmarking. It is clear that the menu of promising approaches will be in constant flux, and new entries need to be encouraged, if they can demonstrate promise.

To broaden participation in this research, agencies can package datasets and challenges together, and publicize these packages. For example, a challenge on determining text structure, or determining affect, could be made available along with appropriate data, and information about procedures for benchmarking. This would ease the path to participation for CS researchers not now working on opportunities in disability.

### 4. Conclusion

As delivery of content and services online increases, the need to support full participation of people with disabilities in the online world also increases, as recognized by international and national priorities. New forms of interaction bring with them the need to make them help include, rather than exclude, people with disabilities. At the same time, the opportunities to build on research progress in other fields are increasing. The time is right for researchers, and those who support research, to step up our efforts to address these needs by grasping these opportunities.

**Workshop Participants**

- Holly Anderson, Department of Education
- Robert Baker, Social Security Administration
- Colin Clark, Fluid Project (Canada)
- Scott Cory, Administration for Community Living
- Boris Goldowsky, CAST, Inc.
- Mark Hakkinen, Educational Testing Service
- Peter Harsha, Computing Research Association
- Shaun Kane, University of Colorado
- Stephen Kell, University of Cambridge
- Raja Kushalnagar, Rochester Institute of Technology
- Ming Lin, Univ. of North Carolina at Chapel Hill
- Sue-Ann Ma, DIAGRAM Center
- Jennifer Mankoff, Carnegie Mellon University
- Jamal Mazrui, Federal Communications Commission
- Margaret Mitchell, Microsoft Research
- Brad Myers, Carnegie Melon University
- Gary O'Brien, IARPA
- Ram (P G) Ramachandran, IBM
- James Rehg, Georgia Institute of Technology
- Luz Rello, Carnegie Melon University
- Bill Schutz, HHS
- Ben Shneiderman, University of Maryland
- Sylvia Spengler, NSF
- Ed Summers, SAS
- John Tschida, HHS
- Christian Vogler, Gallaudet University/Technology Access Program
- Bruce Walker, Georgia Institute of Technology
- Jason White, Educational Testing Service
- Mohammed Yousof, DOT
- Changxi Zheng, Columbia University
- Sina Bahram, NCSU / PrimeAccess consulting
- Antranig Basma, Fluid Project (UK)
- Jeffrey Bigham, Carnegie Mellon University
- Robert Bohn, NIST
- Henry Claypool, ex AAPD
- Sandra Corbett, Computing Research Association
- Khari Douglas, Computing Community Consortium
- Ann Drobnis, Computing Community Consortium
- Amy Hurst, UMBC
- Deborah Kaplan, HHS
- Richard Ladner, University of Washington
- Clayton Lewis, University of Colorado
- Beth Mynatt, Georgia Tech
- Bill Peterson, Department of Homeland Security
- Mike Shebanek, Yahoo
- Gregg Vanderheiden University of Wisconsin
- Helen Wright, Computing Community Consortium

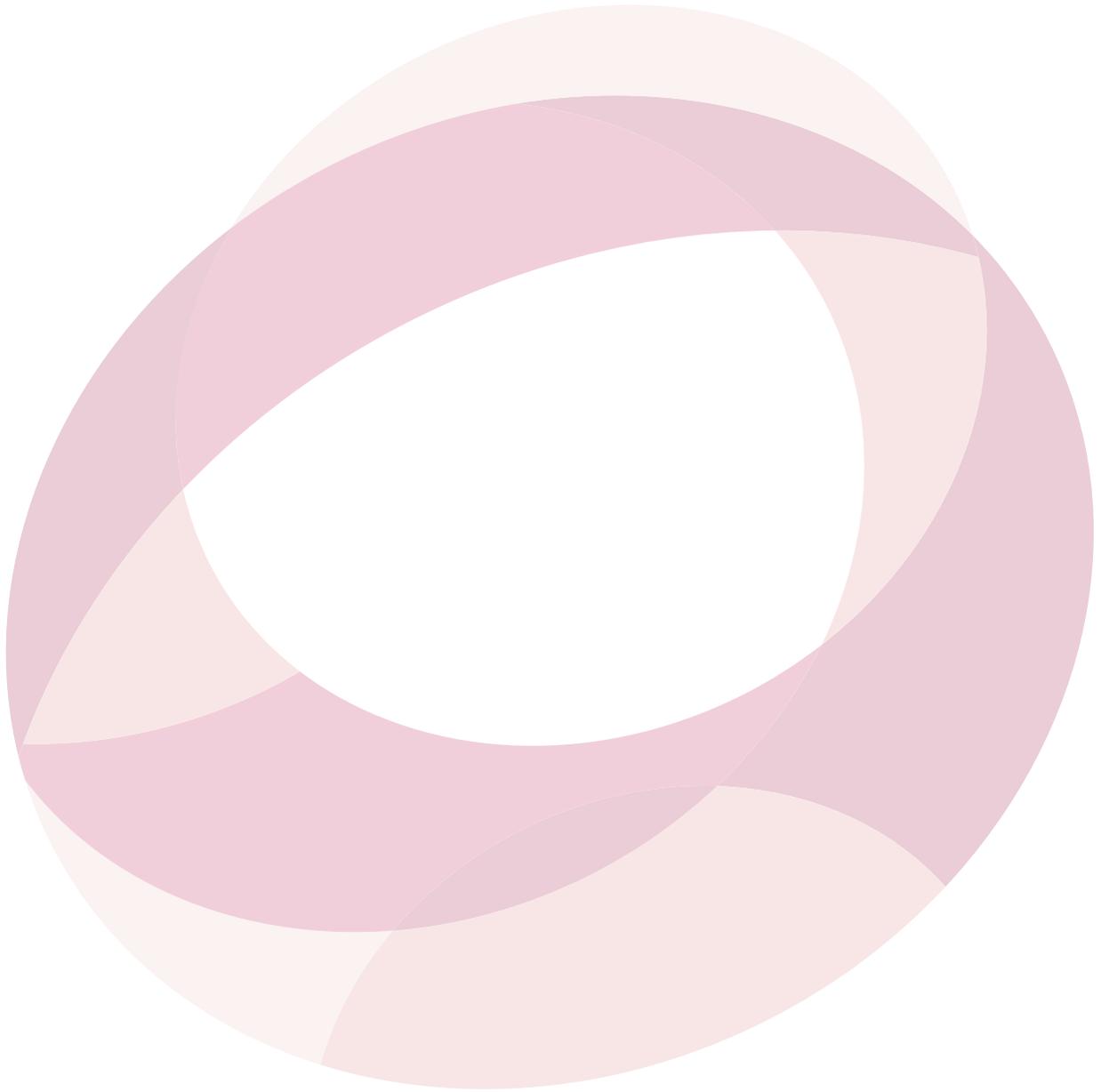

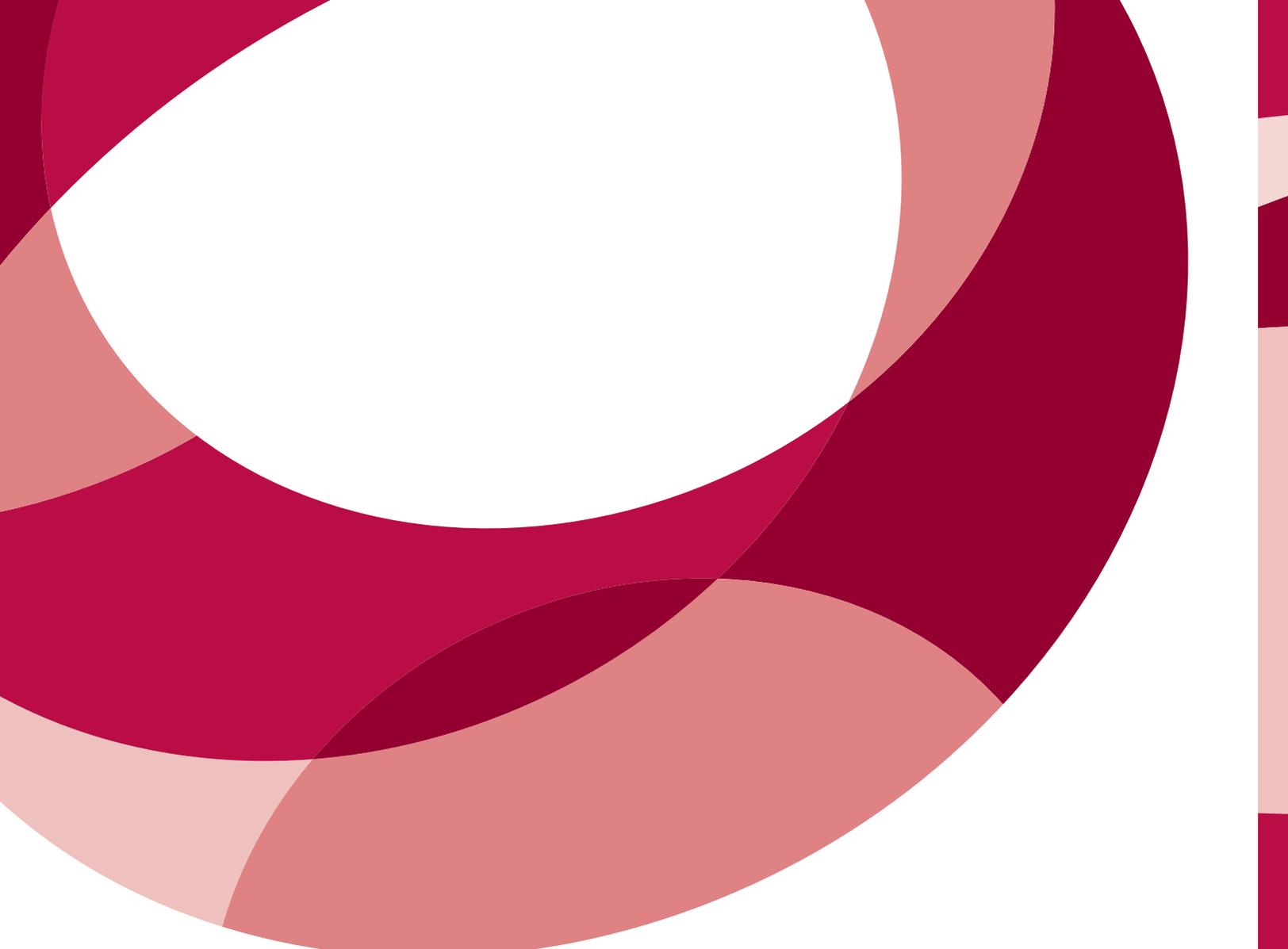

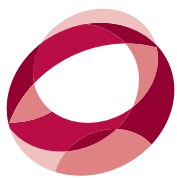
**CCC**
Computing Community Consortium
Catalyst

1828 L Street, NW, Suite 800
Washington, DC 20036
P: 202 234 2111 F: 202 667 1066
www.cra.org cccinfo@cra.org